\documentstyle[12pt]{article}
\begin{document}
\baselineskip 16pt
\newcommand{\be}{\begin{equation}}
\newcommand{\ee}{\end{equation}}
\parskip 10pt
\parindent 25pt
\begin{flushright}{OKHEP 93-004}\end{flushright}
\title{Stochastic Neural Networks with \\ the Weighted Hebb Rule}
\author{Caren Marzban \\
Department of Physics and Astronomy\\
University of Oklahoma, Norman, OK  73019 U.S.A. \\ \\
Raju Viswanathan  \\
Center for Artificial
Intelligence and Robotics \\ Raj Bhavan Circle, Bangalore 560001 India}

\maketitle
\begin{abstract}
Neural networks with synaptic weights constructed according to
the weighted Hebb rule, a variant of the familiar Hebb rule, are
studied in the presence of noise(finite temperature),
when the number of stored patterns
is finite and in the limit that the number of neurons $N\rightarrow
  \infty$. The fact that different patterns enter the synaptic rule
with different weights changes the configuration of the free energy
surface. For a general choice of weights not all of the patterns are
stored as {\sl global} minima of the free energy function. However,
as for the case of the usual Hebb rule, there exists a temperature
range in which only the stored patterns are minima of the free
energy.
In particular, in the presence of a single extra pattern stored with
an appropriate weight in the synaptic rule, the temperature at which
the spurious minima of the free energy are eliminated is significantly
lower than for a similar network without this extra pattern.
The convergence time of the network, together with the overlaps of
the equilibria of the network with the stored patterns, can thereby
be improved considerably.
\end{abstract}
\pagebreak

\section{Introduction}

The statistical mechanics of large neural networks with the Hebb rule
prescription for the synaptic weights has been studied in
detail and is now well--understood[1,2]. In this paper, we shall
study the statistical mechanics of neural nets with synaptic weights
which are constructed according to the weighted Hebb rule. For
orthogonal patterns, the Hebb rule indeed stores the required patterns
as fixed points of the deterministic updating dynamics, as is well known.
The role of the weighted Hebb rule in the storage of non--orthogonal
patterns was examined in ref.[3]. The weighted Hebb rule is also a
natural choice when one wants to store patterns or classes of patterns
with varying radii of attraction. Although the precise dependence of the
radii of attraction on the weights with which different patterns enter
the synaptic rule is difficult to study, it is clear that this rule
offers the possibility of adjusting the radii of attraction.

Our principal motivation for studying the weighted Hebb rule
arises from the expectation that
the presence of different weights for different patterns would affect
the configuration of the free energy surface. There is the possibility
that some of the degeneracy of the minima of the free energy would be
lifted; in addition, the range of useful operating temperatures of the
network would be changed. We shall find that in fact, the critical
operating temperature of the network can be suitably lowered by a
judicious choice of weights. The time needed for the network to
converge to useful equilibrium states can be thereby reduced, since
lower noise levels mean faster convergence times. Additionally, at
lower temperatures, the overlaps of the network equilibria with the
stored memories are larger; the overall quality of memory recall of
the network can thus be significantly enhanced.

In the next section, we present the evaluation of the free energy along
the lines of ref.[1]. The stationary-point conditions yield the mean
field equations (MFE's) for equilibrium states in the large--$N$ limit.
In section 3, we derive the stability conditions of the solutions
to the mean field equations. In section 4, various critical
temperatures for the existence of stable equilibria are calculated.
In the last section we summarise our conclusions.

\section{The weighted Hebb rule}

We start with a network of $N$ neurons with states $s_i(t)=\pm 1$
at time $t$. At time $t+1$, the
probability of $s_i$ flipping sign is
\[ W(s_i\rightarrow -s_i)=(1+{\rm{exp}}(2\beta s_i h_i))^{-1},  \]
where
\[ h_i=\sum_{j=1}^{N}\;J_{ij}s_j(t) \]
is the local field or potential at neuron $i$ due to all the other
neurons, and $\beta$ is an inverse noise parameter(equivalently
$T= 1/{\beta}$ is a `temperature' parameter).

Since $w_{ij}$ is symmetric, an energy function
\[ H=-\frac{1}{2} \sum_{i\neq j}\;J_{ij}s_is_j \]
can be attached to every configuration [$s$]. At zero temperature
($\beta\rightarrow\infty$), the network converges to a local minimum of
this energy function.

Given a set of $p$ patterns(finite in number) $\sigma^{\mu}_i$, $\mu=1,..,p$,
to be stored,
we could try to store these patterns by constructing the
synaptic weights $J_{ij}$ in
the form
\[
J_{ij}=\frac{1}{N}\sum_{\mu, \nu}\;g_{\mu \nu} \sigma_i^{\mu} \sigma_j^{\nu}
\]
where the numbers $g_{\mu\nu}$ are positive. Without the $g_{\mu \nu}$
factor this would be nothing but the usual Hebb rule.
However,with the $g_{\mu \nu}$, if we require the patterns $\sigma$ to be
stored as fixed points in
the $\beta\rightarrow\infty$ limit, the statistically significant
contributions come only from the $g_{\mu\mu}$ terms in $J_{ij}$. We
shall therefore retain only these diagonal terms $g_{\mu\mu}\equiv g_{\mu}$
in the synaptic rule, yielding the weighted Hebb rule:
\[
J_{ij}=\frac{1}{N}\sum_{\mu}\;g_{\mu} \sigma_i^{\mu} \sigma_j^{\mu}.
\]

At finite temperatures $T$, one needs to look for minima of the free energy
to identify metastable states of the network. Accordingly, we need
to evaluate the partition function
\[ Z=\mbox{Tr}e^{-\beta H}. \]
Proceeding as in ref.[1], and assuming that the stored patterns
are random, we define overlap variables $m_{\mu}\equiv \ll \sigma^{\mu}
<s>\gg$.
In the $N\rightarrow\infty$ limit, the free energy per neuron $f=F/N$
and the stationary point equations we get in the evaluation of $Z$
then take the form
\be
f(\beta)=\frac{1}{2}\sum_{\mu}^{p} \frac{1}{g_{\mu}}m_{\mu}^2 -
\frac{1}{\beta}\ll \mbox{log(2 cosh}\beta \vec{m}.\vec{\sigma})\gg\;,
\ee
and
\be
m_{\mu}=g_{\mu}\ll \sigma^{\mu}\mbox{tanh}\beta \vec{m}.\vec{\sigma}\gg
\ee
respectively.

We shall first look for solutions to the stationary point or mean field
equations at zero temperature.

{\bf \underline{$T=0$}}:

In this limit, the tanh function becomes a sign function, and
$ \log (2 {\rm cosh} y)\rightarrow |y| $ as $|y|\rightarrow\infty$.

If $\vec{m}$ has only one non--zero component
(the Mattis states), for instance $\vec{m}=(m,0...0)$, then $m=g_1,$
(up to an irrelevant sign),
 and $f=-(1/2)g_1$. Of course, for a choice of $\vec{m}=(0,m,0...0)$
one would get $m=g_2$ and $f=-(1/2)g_2$, etc., but without loss of
generality we will confine ourselves to only the former ansatz.
{}From this we can see that the lowest-energy state, and hence a stable state,
is
\[ \vec{m}=\pm g_{max}(1,0...0),\]
with $g_{max}$ being the largest component of $\vec{g}$.
The stability of these
states can also be seen from the MFE and the free energy directly: we have
$f=-(1/2)\sum (1/g_{\mu})m_{\mu}^2$ and
$\sum(1/g_{\mu})m_{\mu}^2 \leq g_{max}$
(this follows from $\ll\mid m\cdot\sigma\mid\gg\leq[\sum(m^{\mu})^2]^{1/2}$),
implying that these Mattis states occupy global minima.

We note, however, that the other Mattis states, corresponding to
$g<g_{max}$, are not {\sl global} minima. Nevertheless, they are
certainly local minima (and hence metastable states) at zero
temperature, and they exist
as stable states for sufficiently low temperatures as well, as we shall see.

Any state which is not a Mattis state will be called a {\sl spurious}
state, as is usual; the Mattis states are the ones desirable for
associative memory purposes.

For symmetric states with $n$ non-zero components of the type
$\vec{m}=m_n(1...1,0...0)$, the MFE's imply
that one must require,
\[ g_1=g_2=...=g_n\equiv g \]
in which case we get
\[ m_n=\pm\frac{g}{n}\ll \mid z_n \mid \gg \]
and
\[ f_n=-\frac{1}{2} \frac{g}{n} m_n^2=-\frac{g}{2n}\ll \mid z_n \mid\gg ^2.
\]
We shall call such states symmetric states corresponding
to $g$.

These equations differ from those obtained in [1] only in the factor of $g$
appearing on the right-hand side, resulting in the same ordering of
the $f_n$'s as that of [1].

We can also consider general states of the form $\vec{m}=(m_1,m_2,m_3,...,
m_n,0,...,0)$, with nonzero $m$'s. However, to reduce technical complexity
we restrict ourselves to the case of $n=3$. Since there are no non-trivial
solutions with $n=2$, as we shall see, this will be
sufficient to establish a definitive conclusion. It is easy to see
that the $T=0$ limit of these
states is $\vec{m}=(1/2)(g_1,g_2,g_3,0,...,0)$, and their stability will be
discussed in the next section.

{\bf \underline{$T\neq 0$}}:

We shall assume that the states we wish to look at start appearing
just below a temperature $T$(which depends on the state); correspondingly,
the overlaps $m$ are small near this temperature, and we can expand
the cosh and tanh functions in a series in $m$, keeping only the first
few terms.
Then our equations become (to the appropriate order)
\be m_{\mu}=g_{\mu}\beta m_{\mu}(1+\frac{2}{3}\beta^2 m_{\mu}^2-
\beta^2 \vec{m}^2) \ee
and
\be f(\beta)=\frac{1}{2}\sum_{\mu}\frac{1}{g_{\mu}}m_{\mu}^2-
\frac{1}{2}\beta\sum_{\mu}m_{\mu}^2+\frac{1}{12}\beta^3
\ll (\vec{m}.\vec{\sigma})^4 \gg -T\log 2 .\ee

We can see that there exists a critical temperature, above which
the only solution is the trivial one $\vec{m}=0$.
For example, for the Mattis state $\vec{m}=(m,0...0)$, the solution is
\[ m^2(g_1)= 3g_1(g_1-T)\]
where, again, the subscript ``1'' can be elevated to a $\mu$, since
a general Mattis state can have its non-zero component in any slot. Also
the associated free energy is
\[ f(\beta)+T\log 2 =-\frac{3}{4g_1}(g_1-T)^2.\]
The critical temperature for the appearance of a Mattis state with
one non-zero component $m^{\mu}$ is therefore $T_c^{\mu}=g_{\mu}$.
For all of the Mattis states to exist as solutions of the stationary point
equations, therefore, the operating temperature of the network must
satisfy $T<g_s$ where $g_s$ is the smallest of the weights $g_{\mu}$.

The symmetric states $\vec{m}=m_n(1...1,0,...0)$, still require that
we have $\vec{g}=(g,...g,g_{n+1},...,g_p)$, with $n$ $g$'s, and to
be a solution $m_n$ must be
\[ m_n^2=\frac{3}{3n-2}g(g-T).\]
The corresponding free energies take the form
\[ f+T\log 2=-\frac{3}{4g}\frac{n}{3n-2}(g-T)^2 \]
so that for given $g$, the $n=1$ state has the least free energy among
the symmetric states corresponding to that $g$.
In the next section we will see that these are in fact unstable above
a certain critical temperature, and so we postpone the discussion of
stability to that section.

For the general asymmetric states, having restricted our attention to
the $n=3$ case (i.e. $\vec{m}=(m_1,m_2,m_3,0,...0)$), we shall show that
these states are also unstable at $T=g_{\mu}$. The stability of these
states will be discussed at length in the next section.

\section{Stability}

The positivity of the eigenvalues of the stability matrix $\partial^2 f/
\partial m_{\mu} \partial m_{\nu}$ assures the stability of the states.
{}From (2) we get
\[ \frac{\partial^2f}{\partial m_{\mu} \partial m_{\nu}}=
\frac{1}{g_{\mu}}\delta_{\mu \nu}-\beta(\delta_{\mu \nu}-Q_{\mu \nu}),\]
where
\[ Q_{\mu \nu}=\ll \sigma^{\mu} \sigma^{\nu} \mbox{tanh}^2\beta \vec{m}.
\vec{\sigma}\gg . \]

{\bf \underline{Zero temperature}:}

As we discussed in the previous section the Mattis states are stable at
$T=0$ for $\vec{g}=(g_{max},g_2,...,g_p)$, with $\vec{m}=(g_{max},0,...,0)$
being the global minimum.

For the symmetric states with $n$ non-zero components
$\vec{m}=m_n(1,1,..,1,0,0,...0)$, with $\vec{g}=(g,...,g,g_{\alpha})$,
  and
$\alpha=n+1,...,p$, we find the eigenvalues of the stability matrix to be
\begin{eqnarray}
\lambda_1 &=& \frac{1}{g}-\beta(1-q_n)+\beta(n-1)Q \nonumber \\
\lambda_2^{(\alpha)} &=& \frac{1}{g_{\alpha}}-\beta(1-q_n) \\
\lambda_3 &=& \frac{1}{g}-\beta(1-q_n)-\beta Q \nonumber
\end{eqnarray}
where $q_n=Q_{\mu \mu}$.

As in ref.[1], in the $T\rightarrow 0$ limit, the parameter $q$ stays
finite for even $n$, and goes to unity exponentially in $\beta$ for odd $n$,
 while $Q$ goes to zero exponentially. Therefore the eigenvalues are all
positive for the odd $n$ states, while the even $n$ states are all unstable
due to the presence of negative eigenvalues.

In the case $n=3$, for instance, and in the limit $T=0$,
we see that $q_n=1$ and $Q=0$, giving $\lambda_1=1/g$, $\lambda_2^{\alpha}
=1/g_{\alpha}$, and $\lambda_3=1/g$, all of which are positive, yielding
stability.

Similarly, for $T=0$ and for asymmetric states, all the $p$
eigenvalues reduce to their
respective $1/g_{\mu}$, again yielding stability.

{\bf \underline{Finite temperatures}:}

For the symmetric states, at the temperature $T\sim g$, we have
$q\approx\beta^2 m_n^2 n$, and $Q=(2q/n)$. Then we can see that
\[ \lambda_3=\frac{1}{g}-\beta(1-q)-\beta Q\approx
-\frac{4}{3n-2}\frac{1}{g^2}(T-g) \]
which is clearly negative for $T\leq g$,
except for $n=1$ where $\lambda=\frac{2\beta}{g_{\mu}}(g_{\mu}-T)>0$.
The ($n>1$) symmetric states are therefore unstable at $T=g$.

Let us mention in passing that $\lambda_1 = \frac{1}{g}-\beta(1-q)+
\beta(n-1)Q$ becomes, for $n=1$ states, $+\frac{2(g-T)}{g^2}$,
which is positive below the temperature $T=g$. The
eigenvalue $\lambda_3$ is not present for $n=1$ states. The sign
of $\lambda_2$ depends explicitly on the various components of $\vec{g}$,
and this shall be discussed further, below. However, that $\lambda_3$ is
negative is sufficient to render the symmetric states with $n>1$
unstable. The exact
temperature at which their stability, as well as the stability of the
asymmetric states, is lost will be derived in the next section.

\section{Critical Temperatures}

First, we deal with the symmetric states.  We showed that of the eigenvalues
(5) of the stability matrix, $\lambda_1$ is positive in the range $T=0$ to
$T\sim g$, $\lambda_3$ changes sign from $+$ to $-$, while the sign of
$\lambda_2$ depends on the form of $\vec{g}$ explicitly (see below).
Hence, there are two possibilities to consider: one is where $\lambda_3$
is set to zero, to find the critical temperature $T=T_c$ at which
$\lambda_3$ changes sign, while $\lambda_2$
is constrained to be positive at that temperature $T_c$. The second
case is where
$\lambda_2$ is set to zero to find the critical temperature $T=T_c^{\ast}$,
at which $\lambda_2$ changes sign, while $\lambda_3$ is constrained
to be positive. The former gives
\be T_c=g(1-q_n+Q) \ee
and requiring $\lambda_2 \geq 0$ at $T=T_c$ gives the constraint
\be \frac{g_{\alpha}}{g}\leq \frac{1-q_n+Q}{1-q_n} \ee
where we recall $\vec{g}=(g,...,g,g_{\alpha})$, with $n$ components
equal to $g$ and
$\alpha=n+1,...,p$.
The MFE (2), when specialized to $n=3$ symmetric states yields
\be x=\frac{1}{4(T/g)}(\tanh x+\tanh 3x),\ee
where $x=\beta m_3$. With $q_3=(1/4)(\tanh^23x+3\tanh^2x)$ and
$Q=(1/4)(\tanh^23x-\tanh^2x)$, solving (6) and (8) numerically for $x$ and
$T_c/g$, we obtain
\[ x=0.94,\;\;\;\;\frac{T_c}{g}=0.46,\;\;\;\;\mbox{for} \;\;\;\;
\frac{g_{\alpha}}{g} \leq 1.32 \]
with the last constraint coming from (7).

Our results up to this point do not differ significantly from those
of [1]. However, let us go on further to the second case with
$n=1$.

Let $g_s$ be the smallest of the $g$'s, and consider the
corresponding Mattis state $\vec{M}=(0,0,..,m_s,0,..,0)$. The smallest of the
eigenvalues in this case is $\lambda_2^{\alpha}$ with $g_{\alpha}=g_{max}$,
\[
\lambda^{\alpha}_2=\frac{1}{g_{max}}-\beta(1-q^s_1).
\]
where $g_{max}$ is the largest of the $g$'s, and $q_1^s$ is the corresponding
value of $q$. Now to avoid spurious
$n=3$ states corresponding to $g_{max}$ (which exist whenever $g_{max}$
occurs at least three times in the set of $g$'s), the operating
temperature of the network must be greater than $T_c=0.46 g_{max}$, as
we have seen. At this temperature, in order for $\vec{M}$ to exist as
a stable state, $\lambda^{\alpha}_2$ must be positive, or at best  zero.
This gives the relation
\[
1-q_1^s<0.46;
\]
together with the MFE
\[ \frac{m_s}{g_s}={\rm{tanh}}\beta m_s, \]
this yields the constraint $g_s/g_{max}>0.589$ on the value that the smallest
$g$ can take, if {\underline{all}} the given patterns
are to be stored as stable Mattis states of the network.

Turning now to the case of the $n=3$ symmetric
states corresponding to $g$, and for $g_{\alpha}/g> 1.32$,
where $g_{\alpha}$ occurs only once or two times among the
$g$'s, we see that
\[ \frac{T_c^{\ast}}{g}=\frac{g_{\alpha}}{g}(1-q_n),\;\;\mbox{with}\;\;
\frac{g_{\alpha}}{g} \geq 1.32  \]
some of whose solutions can be tabulated as follows:

\begin{center}
\begin{tabular}{|l|l|l|l|l|l|r|}  \hline
$g_{\alpha}/g$  & 1.32  & 1.34 & 1.42 & 1.66 & 2.0 & 3.0   \\ \hline
$x$             & 0.94  & 0.96 & 1.04 & 1.21 & 1.37& 1.69  \\ \hline
$T_c^{\ast}/g$  & 0.46  & 0.45 & 0.43 & 0.38 & 0.34& 0.29  \\ \hline
\end{tabular}
\end{center}

We can now see that, whereas for $g_{\alpha}\leq 1.32 g$, the critical
temperature is
simply $0.46 g$, for $g_{\alpha}\geq 1.32 g$, the critical temperatures
are all lower than the former. If $g_0$ is the largest $g$ which occurs
at least three times, the operating temperature of the network must
be at least $0.46 g_0$ if the largest
$g$ bigger than $g_0$, $g_{max}$,
satisfies $g_{max}\leq 1.32 g_0$. This minimum necessary
temperature for the avoidance of spurious equilibria is
lowered when $g_{max}>1.32 g_0$. In other words, by adding additional
patterns with sufficiently large weights, we can lower the temperature
above which there are no spurious states, leading to a ``better" network.
What is meant by ``better" will be discussed in the next section.

The symmetric states with $n >3$ can be shown to have even lower
critical temperatures, exactly as in [1]. Therefore, it is sufficient
to consider the $n=3$ states only.

We now proceed to the case of the asymmetric states $\vec{m}=(m_1,m_2,m_3,
0,...,0)$ with a general weight vector, i.e.
$\vec{g}=(g_1,g_2,g_3,g_{\alpha})$. The MFE's can be written in the form
\begin{eqnarray*}
x+y &=& \frac{1}{2} \frac{1}{(T/g_1)}
[\tanh(x+y+z)+\tanh x+\tanh y-\tanh z] \nonumber \\
x+z &=& \frac{1}{2} \frac{(g_2/g_1)}{(T/g_1)}
[\tanh(x+y+z)+\tanh x-\tanh y+\tanh z] \\
y+z &=& \frac{1}{2} \frac{(g_3/g_1)}{(T/g_1)}
[\tanh(x+y+z)-\tanh x+\tanh y+\tanh z] \nonumber
\end{eqnarray*}
where $x=\beta(m_1+m_2-m_3),\;y=\beta(m_1-m_2+m_3),\;z=\beta(-m_1+m_2+m_3)$,
and
the secular equation, dictating stability, can be written as
\[ [\lambda^3+\frac{1}{T}l_2\lambda^2+\frac{1}{T^2}l_1\lambda
+\frac{1}{T^3}l_0]\prod_{\alpha=n+1}^{p}[\frac{1}{g_{\alpha}}-
\beta(1-q)-\lambda]=0 \]
where
\begin{eqnarray*}
l_2 &=& 3(1-q)-(T/g_1)\left(1+\frac{1}{(g_2/g_1)}+\frac{1}{(g_3/g_1)}\right) \\
\rule{0mm}{7mm}
l_1 &=& 3(1-q)^2-Q_1^2-Q_2^2-Q_3^2 \nonumber   \\
    &-& 2(T/g_1)(1-q)
\left(1+\frac{1}{(g_2/g_1)}+\frac{1}{(g_3/g_1)}\right) \nonumber  \\
    &+& (T/g_1)^2\left(\frac{1}
{(g_2/g_1)}+\frac{1}{(g_3/g_1)}+\frac{1}{(g_2/g_1)(g_3/g_1)}\right) \\
\rule{0mm}{10mm}
l_0 &=& -\left(2Q_1Q_2Q_3+(1-q)(Q_1^2+Q_2^2+Q_3^2)-(1-q)^3\right) \nonumber \\
    &+& (T/g_1)\left(\frac{Q_1^2}{(g_3/g_1)}+\frac{Q_2^2}{(g_2/g_1)}+Q_3^2
-(1-q)^2(1+\frac{1}{(g_2/g_1)}+\frac{1}{(g_3/g_1)})\right) \nonumber \\
    &+& (T/g_1)^2(1-q)\left(\frac{1}{(g_2/g_1)}+\frac{1}{(g_3/g_1)}+
\frac{1}{(g_2/g_1)(g_3/g_1)}\right)\nonumber  \\
    &-& \frac{(T/g_1)^3}{(g_2/g_1)(g_3/g_1)}
\end{eqnarray*}
and
\begin{eqnarray*}
Q_1 \equiv Q_{12} &=& \frac{1}{4}
[\tanh^2(x+y+z)+\tanh^2x-\tanh^2y-\tanh^2z] \\
Q_2 \equiv Q_{13} &=& \frac{1}{4}
[\tanh^2(x+y+z)-\tanh^2x+\tanh^2y-\tanh^2z] \\
Q_3 \equiv Q_{23} &=& \frac{1}{4}
[\tanh^2(x+y+z)-\tanh^2x-\tanh^2y+\tanh^2z] \\
q &=& \frac{1}{4}
[\tanh^2(x+y+z)+\tanh^2x+\tanh^2y+\tanh^2z]
\end{eqnarray*}

Since we are generally interested in the temperature $T_c$ at which a given
eigenvalue becomes zero (i.e. changes sign from $+$ to $-$), there are
two separate cases we can consider: one is where $\lambda^{(\alpha)}
=\frac{1}{g_{\alpha}}-\beta(1-q)$ is set to zero, while the other 3
eigenvalues
(from the cubic part) are constrained to be nonnegative. The second choice is
to set one of the 3 eigenvalues from the cubic part equal to
zero and demand for
$\lambda^{(\alpha)}$ and the remaining 2 eigenvalues to be non-negative.

The former case gives
\be  (T_c/g_1)=\frac{g_{\alpha}}{g_1}(1-q) \ee
and the positivity of the other eigenvalues can be insured by
the constraints
\[ l_2 < 0, \;\; l_1 > 0, \;\; \mbox{and} \;\; l_0 < 0. \]
The first of these constraints, in conjunction with (9), simplifies to
\[ \frac{g_{\alpha}}{g_1} > 3\left(1+\frac{1}{(g_2/g_1)}
+\frac{1}{(g_3/g_1)}\right)^{-1} \]
The last two constraints, due to their dependance on the $Q$'s and the $q$,
must be imposed numerically in finding $T_c$. Some results are shown
in the table below for the case when $g_2=g_3$. These spurious
states are {\sl stable} for $g_{\alpha}>g_{\alpha}^{min}$ and $T<T_c^{\ast}$.

\begin{center}
\begin{tabular}{|l|l|l|l|r|}   \hline
$g_2/g_1$  & 0.6 & 0.8 & 0.95 & 1.0  \\  \hline
$g_{\alpha}^{min}/g_1$  & 2.0 & 1.89 & 1.6 & 1.32  \\ \hline
$T_c^{\ast}/g_1$  & 0.18 & 0.27 & 0.37 & 0.46 \\ \hline
\end{tabular}
\end{center}

In the second case, since we are interested only in the zero eigenvalues,
it is sufficient to set $l_0 = 0$, and solve this equation numerically along
with the MFE's. For the remaining eigenvalues to be nonnegative we must
require
\[ l_2 < 0, \;\; l_1 > 0, \;\; {\mbox{and}} \;\; \frac{g_{\alpha}}{g_1} <
\frac{(T/g_1)}{1-q} \]
Some results of this calculation are shown in the table below for $g_3=g_2$.

\begin{center}
\begin{tabular}{|l|l|l|l|l|r|}  \hline
$g_2/g_1$  & 1.1 & 1.2 & 1.3 & 1.4 & 1.5 \\  \hline
$T_c/g_1$  & 0.29 & 0.22 & 0.19 & 0.15 & 0.11 \\ \hline
\end{tabular}
\end{center}

Again, these spurious states are {\sl stable} for $T<T_c$.

We note that for $\vec{g}$ of the form $(g_1,g_2,g_3,...)$, with $g_1=g_2=1$
and $g_3=1.32$, the critical temperature of the associated spurious state
$(m_1,m_2,m_3,0..,0)$ with $m_1=m_2$ is close to 0.19. If $g_1$ occurs at
least three times in $\vec{g}$, the critical temperature of the $n=3$
symmetric state corresponding to $g_1$ is 0.46. We can in fact make the
general statement that if $g_{max}$ is the largest component of $\vec{g}$,
and $g_0$ the second largest, for $g_{max}/g_0>1.32$, the critical
temperature above which there are {\sl no} spurious states is determined
by demanding the instability of spurious states with non-zero entries
$m_i$ of $\vec{m}$ corresponding to $g_i\leq g_0$.

A set of results for asymmetric states with $g_2\neq g_3$ are also
given in the following table.
\begin{center}
\begin{tabular}{|l|l|l|}  \hline
$g_2/g_1$  & $g_3/g_1$ & $T_c/g_1$ \\  \hline
0.93 & 0.55 & 0.12  \\ \hline
0.91 & 0.82 & 0.21 \\ \hline
0.9 & 0.6 & 0.13 \\ \hline
0.85 & 0.7 & 0.15 \\ \hline
0.8 & 0.7 & 0.13 \\ \hline
0.8 & 0.6 & 0.10 \\ \hline
0.75 & 0.65 & 0.10 \\ \hline
0.7 & 0.66 & 0.09 \\ \hline
\end{tabular}
\end{center}
We can also present our results in the following format that clarifies
the behaviour of $T_c/g_1$ for various values of $g_2/g_1$ and $g_3/g_1$:
\begin{center}
\begin{tabular}{l|l|l|l|l|l|}
$\frac{g_3}{g_1}$$\backslash$$\frac{g_2}{g_1}$
& 0.8 & 0.9 & 1.0 & 1.1 & 1.2 \\ \hline
0.8 & .12 &     & .27 &     & .15 \\ \hline
0.9 &     & .25 & .33 & .23 &     \\ \hline
1.0 & .27 & .33 & .46 & .29 & .22 \\ \hline
1.1 &     & .23 & .29 & .37 &     \\ \hline
1.2 & .15 &     & .22 &     & .35 \\ \hline
\end{tabular}
\end{center}
The apparent symmetry of this table is simply due to the symmetry of the
MFE's and the secular equation under the simultaneous exchange of
$2\leftrightarrow 3$ and $x\leftrightarrow y$. It is now evident that all
the critical temperatures we have
obtained for the asymmetric states are smaller than $0.46g_0$ (where
$g_0$ is the largest $g$ that occurs atleast three times), as one
moves away from the Hebbian case at the center of the table.

\section{Conclusion}

Our investigation of the use of the weighted Hebb rule in Hopfield
networks has revealed that the structure of the minima of the free
energy at finite temperatures can be quite distinct from the case
of the usual Hebb rule. In particular, by choosing the weighting
factors for the various patterns appropriately, spurious
states can be destabilised at a
significantly lower temperature compared to that for the usual Hebb rule.
When the operating temperature of the network is larger than the
largest among the critical temperatures for the
various spurious states, we
have a network where only the Mattis states (corresponding to the
stored patterns) are equilibria of the network.

Specifically, we can make the following rather general statements.

(1) If the largest of the $g$'s, $g_{max}$, occurs at least three times
or more, then the temperature range in which no spurious states
exist is $0.46g_{max}<T<g_{max}$. If the largest $g$ which occurs
at least three times is $g_0$, and the largest $g$, $g_{max}$,
occurs no more than two times and satisfies $g_{max}>1.32g_0$, then
the critical temperature above which no stable spurious states exist
is smaller than $0.46g_0$, and can be calculated as we have shown.

(2) If the smallest of the $g$'s is $g_{min}$, and the largest one,
$g_{max}$, occurs at least three times, and
the constraint $g_{min}/g_{max}>0.589$, is satisfied, {\underline{all}}
of the patterns
to be stored exist as stable Mattis states in the range of temperatures
where spurious states are excluded. If $g_{max}$ occurs
only once or twice, this constraint on the ratio
of $g_{min}$ to $g_{max}$ is changed
and can be calculated in a manner analogous to that shown in section 4.

One consequence of the lowering of the useful operating temperature is
that convergence of the network to metastable states would be faster.
A second consequence is that the overlaps of the equilibria of the network
with the stored patterns would be larger due to the reduced temperature.
Given a set of patterns to be stored, one could then simply put in an
extra pattern weighted by a sufficiently larger weight $g$ as compared to
the $g$'s of the other patterns to construct the synapses. The resulting
network would then converge to an equilibrium state closer to one of the
stored patterns and at a faster rate than
a network constructed without this extra pattern being taken into
account. It would be interesting to carry out detailed simulations of
networks employing the weighted Hebb rule and to determine the
relative sizes of the basins of attraction for the different stored
patterns.

{\bf Acknowledgements}

C.M. is grateful to John Keuhler for assistance in the use of Fortran.

\end{document}